\documentstyle[aps,psfig]{revtex}

\begin{document}

\twocolumn[\hsize\textwidth\columnwidth\hsize\csname
@twocolumnfalse\endcsname

\title{Theory of Spike Spiral Waves in a Reaction-Diffusion System}

\author{C. B. Muratov}

\address{Courant Institute of Mathematical Sciences, New York
  University, \\ 251 Mercer St., New York, NY 10012}

\author{V. V. Osipov}

\address{Department of Theoretical Physics, Russian Science Center
  ``Orion'', \\ Plekhanova St. 2/46, 111123 Moscow, Russia}

\date{\today}

\draft

\maketitle

\begin{abstract}

We discovered a new type of spiral wave solutions in
reaction-diffusion systems --- spike spiral wave, which significantly
differs from spiral waves observed in FitzHugh-Nagumo-type models. We
present an asymptotic theory of these waves in Gray-Scott model. We
derive the kinematic relations describing the shape of this spiral and
find the dependence of its main parameters on the control
parameters. The theory does not rely on the specific features of
Gray-Scott model and thus is expected to be applicable to a broad
range of reaction-diffusion systems.

\end{abstract}

\pacs{PACS Number(s): 03.40.Kf, 05.70.Ln, 82.20.Mj, 82.40.Bj}

\vskip2pc]

\bibliographystyle{prsty}

Formation of rotating spiral waves (rotors) is one of the most vivid
and ubiquitous phenomena of nonlinear physics
\cite{winfree,mikhailov,vasiliev,murray,field,cross93,kapral,%
tyson88,ko:book,ko:ufn89,chaos98}. These waves are observed in
nonlinear optical media \cite{akhmanov89}, chemical reactions of
Belousov-Zhabotinsky type, catalytic reactions on crystal surfaces
\cite{field,cross93,kapral}, and in a variety of biological systems:
social amoebae Dictyostelium discoideum \cite{siegert89}, Xenopus
oocytes \cite{lechleiter91}, chicken retina \cite{gorelova83}, and the
heart of animals and man, where the formation of spiral waves is
responsible for cardiac arrythmias and the life-threatening condition
of ventricular fibrillation \cite{winfree,chaos98}.

A generic model used to describe spiral waves is a pair of
reaction-diffusion equations of activator-inhibitor type
\cite{winfree,mikhailov,vasiliev,murray,field,cross93,kapral,%
tyson88,ko:book,ko:ufn89}
\begin{eqnarray}
\tau_\theta {\partial \theta \over \partial t} & = & l^2 \Delta \theta
- q(\theta, \eta, A), \label{act:gen} \\ \tau_\eta {\partial \eta
\over \partial t} & = & L^2 \Delta \eta - Q(\theta, \eta, A),
\label{inh:gen} 
\end{eqnarray}
where $\theta$ is the activator, i.e., the variable with respect to
which there is a positive feedback; $\eta$ is the inhibitor, i.e., the
variable with respect to which there is a negative feedback and which
controls activator's growth; $q$ and $Q$ are certain nonlinear
functions representing the activation and the inhibition processes;
$l$ and $L$ are the characteristic length scales, and $\tau_\theta$
and $\tau_\eta$ are the characteristic time scales of the activator
and the inhibitor, respectively; and $A$ is the bifurcation
parameter. A considerable amount of studies was done on the excitable
systems with FitzHugh-Nagumo-type kinetics (N-systems) (see, for
example,
\cite{winfree,mikhailov,vasiliev,murray,field,cross93,kapral,tyson88}
and references therein). These systems are described by
Eqs. (\ref{act:gen}) and (\ref{inh:gen}) with $L = 0$, and the
nonlinearity in $q$ such that the nullcline of Eq. (\ref{act:gen}) is
N- or inverted N-shaped. Theory of the spiral waves in N-systems with
$\alpha = \tau_\theta / \tau_\eta \ll 1$ was recently developed by
Karma \cite{karma91,karma92}.

The existence of spiral waves in excitable N-systems is due to the
ability of such systems to sustain traveling waves, the simplest of
which is a solitary wave --- traveling autosoliton (AS)
\cite{winfree,mikhailov,vasiliev,murray,field,cross93,kapral,tyson88,%
ko:book,ko:ufn89}. In N-systems the equation $q(\theta, \eta, A) = 0$
has three roots: $\theta_{i1}$, $\theta_{i2}$, and $\theta_{i3}$, for
fixed $A$ and $\eta = \eta_i$. For $\alpha \ll 1$ an AS consists of a
front, which is a wave of switching from the stable homogeneous state
$\theta = \theta_h$ and $\eta = \eta_h$ to the state with $\theta =
\theta_{\mathrm max}$ ($\theta_h = \theta_{i1}$ and $\theta_{\mathrm
max} = \theta_{i3}$ for $\eta_i = \eta_h$) whose width is of order
$l$, and a back of width of order $l$ that follows the front some
distance $w \gg l$ behind the front. Thus, in the AS the distribution
of $\theta$ is a broad pulse, while the value of $\eta$ slowly varies
from $\eta = \eta_h$ to some value $\eta = \eta_m$ in the back of the
pulse, and then slowly recovers from $\eta_m$ to $\eta_h$ behind
it. In the limit $\alpha \rightarrow 0$ the amplitude of the wave (the
value of $\theta_{\mathrm max}$) becomes independent of $\alpha$ and
the speed $c$ does not exceed the value of order $l / \tau_\theta$,
with both $\theta_{\mathrm max}$ and $c$ determined only by the
nonlinearity in $q$.

At the same time, many excitable systems are described by
Eqs. (\ref{act:gen}) and (\ref{inh:gen}) in which the nullcline of
Eq. (\ref{act:gen}) is $\Lambda$- or V-shaped. Examples of such
$\Lambda$-systems are the well-known Brusselator and Gray-Scott models
of autocatalytic reactions and an example of a V-system is the
Gierer-Meinhardt model of morphogenesis \cite{ko:book}. Recently, we
showed for the excitable Brusselator \cite{om:prl95} and Gray-Scott
model \cite{mo1:98} that they are also capable of propagating
traveling waves --- traveling spike AS. The properties of these AS are
essentially different from those in N-systems. The distribution of
$\theta$ in an AS has the form of a narrow spike whose amplitude grows
as $\alpha$ decreases and can become huge as $\alpha \rightarrow
0$. In contrast to N-systems, in the spike $\eta$ varies abruptly and
then slowly recovers back to $\eta_h$ far behind the spike. The speed
of such an AS is always greater than $l / \tau_\theta$ and goes to
infinity as $\alpha \rightarrow 0$. Also, it is important to emphasize
that in an AS the front and the back are not separated by a large
distance, as in N-systems. In this Letter we will show that in
excitable $\Lambda$- or V-systems it may also be possible to excite
steadily rotating spiral waves and develop a theory of such waves in
the case $\alpha \ll 1$.

To be specific, we will consider excitable ($L = 0$) Gray-Scott model,
which is described by the following equations \cite{gray}
\begin{eqnarray}
{\partial \theta \over \partial t} & = & \Delta \theta + A \theta^2
\eta - \theta, \label{act} \\
\alpha^{-1} {\partial \eta \over \partial t} & = & - \theta^2 \eta + 1
- \eta, \label{inh}
\end{eqnarray}
where we chose $l$ and $\tau_\theta$ as the units of time and space,
respectively. Recently, we showed numerically that a steadily rotating
spiral waves may be excited in this model at sufficiently small
$\alpha$ \cite{mo1:98}. From these simulations one can see that the
spiral wave is a steadily rotating slightly curved sharp spike front
of width of order 1 which in the cross-section looks like a periodic
traveling wave train (sufficiently far from the core). We would like
to emphasize that it is this kind of the concentration profiles that
is experimentally observed in Belousov-Zhabotinsky reaction
\cite{skinner91,muller95}.

Let us derive the equation of motion for the traveling sharp front in
Gray-Scott model with $\alpha \ll 1$. Since $\eta$ varies slowly
outside the front, it can be replaced by a constant value $\eta =
\eta_i$ ahead of the front. Let us introduce self-similar variable $z
= \rho - c t$, where $\rho$ is the coordinate along the normal
direction to the front. For definiteness we will assume that $c > 0$,
what means that the front is moving in the $+z$ direction. In the
sharp front the value of $\theta$ is large \cite{mo1:98}, so one can
neglect the last two terms in Eq. (\ref{inh}). Therefore, in the
presence of small curvature $K$ Eqs. (\ref{act}) and (\ref{inh}) can
be written as
\begin{eqnarray} 
{d^2 \theta \over d z^2} + (c + K) {d \theta \over d z} + A \theta^2
\eta - \theta & = & 0, \label{trava} \\ \alpha^{-1} c {d \eta \over d
z} - \theta^2 \eta & = & 0. \label{travi}
\end{eqnarray}
These equations have to be supplemented by the boundary conditions
$\theta(\pm \infty) = 0$ and $\eta(+\infty) = \eta_i$, where the
infinity actually means sufficiently far ahead of the front compared
to the front thickness.

Let us introduce new variables
\begin{equation} \label{new}
\tilde\theta = \alpha^{1/2} \theta \left( {c + K \over c}
\right)^{1/2}, ~~~\tilde\eta = {\eta \over \eta_i},
\end{equation}
and the following quantities
\begin{equation} \label{norm}
\tilde{A} = A \alpha^{-1/2} \eta_i\left( {c \over c + K}
\right)^{1/2}, ~~~\tilde{c} = c + K.
\end{equation}
In these variables Eqs. (\ref{trava}) and (\ref{travi}) become
\begin{eqnarray} 
{d^2 \tilde\theta \over d z^2} + \tilde{c} {d \tilde\theta \over d z}
+ \tilde{A} \tilde\theta^2 \tilde\eta - \tilde\theta & = & 0,
\label{travtheta} \\ \tilde{c} {d \tilde\eta \over d z} - \tilde\theta^2
\tilde\eta & = & 0, \label{traveta}
\end{eqnarray}
with the boundary condition $\tilde\eta(+\infty) = 1$. Observe that
all $\alpha$-, $A$- and $\eta_i$-dependences now enter only via the
parameter $\tilde{A}$. 

\begin{figure} 
\centerline{\psfig{figure=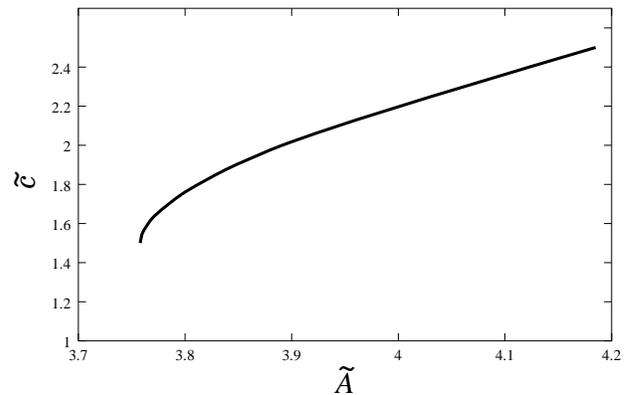,width=8cm}}
\caption{Dependence $\tilde{c}(\tilde{A})$ obtained from the numerical
solution of Eqs. (\ref{travtheta}) and (\ref{traveta}).}
\label{f1}
\end{figure}

Equations (\ref{travtheta}) and (\ref{traveta}) have been analyzed by
us in \cite{mo1:98}. There we showed that these equations indeed admit
solutions in the form of a traveling spike. In the spike $\tilde\eta$
varies from $\tilde\eta = 1$ at $z = +\infty$ (ahead of the front) to
some $\tilde\eta = \tilde\eta_{\mathrm min}$ at $z = -\infty$ (behind
the front). The value of $\tilde\theta \sim 1$ in the spike, so in the
original variables we indeed have $\theta \sim \alpha^{-1/2} \gg 1$
[see Eq. (\ref{new})]. The speed of the front $\tilde{c}$ as a
function of $\tilde{A}$ obtained from the numerical solution of
Eqs. (\ref{travtheta}) and (\ref{traveta}) is presented in
Fig. \ref{f1} \cite{mo1:98}. One can see that this solution exists
only for $\tilde{A} > \tilde{A}_b$, and its speed is always greater
than $c = c_{\mathrm min}$, where
\begin{equation} \label{ab}
\tilde{A}_b = 3.76, ~~~c_{\mathrm min} = 1.5
\end{equation}
The analysis of Eqs. (\ref{travtheta}) and (\ref{traveta}) also shows
that $\tilde\eta_{\mathrm min} = \tilde\eta_{\mathrm min}^b = 0.05$ at
$\tilde{A} = \tilde{A}_b$ and rapidly decreases as $\tilde{A}$
increases \cite{mo1:98}. 

From Eq. (\ref{norm}) immediately follows that for small $K$ the
correction $\delta c$ to the velocity $c$ due to curvature is
\begin{equation} \label{dc}
\delta c = - K \left( 1 + {\tilde{A} \over 2 \tilde{c}} { d \tilde{c}
\over d \tilde{A}} \right),
\end{equation}
where in the right-hand side $\tilde{A}$, $\tilde{c}$, and $d
\tilde{c} / d \tilde{A}$ are evaluated at $K = 0$.  Also, as we showed
in \cite{mo1:98}, for $\tilde{A}$ not in the immediate vicinity of
$\tilde{A}_b$ with good accuracy $\tilde{c} = 0.86 \tilde{A}$ and
$\tilde\eta_{\mathrm min} = 0$. Then, going back to the original
variables, we may write
\begin{equation} \label{ci}
c = c_\infty - a K,
\end{equation}
where 
\begin{equation} \label{ca}
c_\infty = 0.86 A \alpha^{-1/2} \eta_i, ~~~a = {3 \over 2}.
\end{equation}

Behind the sharp front the value of $\eta$ drops from $\eta_i$ to
$\eta_{\mathrm min}$ and $\theta$ goes exponentially to zero
\cite{mo1:98}. On the much longer time scale $\alpha^{-1}$ the value
of $\eta$ recovers from $\eta_{\mathrm min}$ according to
Eq. (\ref{inh}), in which outside of the front the term $-\theta^2
\eta$ can be neglected. From this we obtain that after the front
passed a point $x$ at time $t_i = t_i(x)$ we have
\begin{equation} \label{slow}
\eta(x, t) = 1 - [1 - \eta_{\mathrm min} (t_i)] e^{- \alpha (t -
t_i)},
\end{equation}
where $\eta_{\mathrm min} = \eta_i \tilde\eta_{\mathrm min}$. In a
steadily rotating spiral wave we must have $\eta(x, t_i + T) =
\eta_i(t_i) = {\mathrm const}$, where $T = 2 \pi/ \omega$ and $\omega$
is the angular frequency of the rotation of the spiral. Therefore, the
spiral should be described by Eq. (\ref{ci}) with $c_\infty = \mathrm
const$, which is in turn related to $\omega$. This equation was first
analyzed by Burton, Cabrera, and Frank (BCF) for growth of screw
dislocations on crystal surfaces \cite{burton51} (see also
\cite{tyson88}). They calculated the shape of the spiral and its
frequency in the case when the tip of the spiral is at rest. Applying
their results to Eq. (\ref{ca}), we obtain $\omega = 0.16 \alpha^{-1}
A^2 \eta_i^2$, where for simplicity we used the expressions in
Eq. (\ref{ca}) and put $\eta_{\mathrm min} = 0$. Since $A \eta_i
\gtrsim \alpha^{1/2}$ in order for the front to be able to propagate,
we must have $\omega \gtrsim 1$, so one can expand the exponential in
Eq. (\ref{slow}), and obtain $\eta_i = 3.4 \alpha^{2/3} A^{-2/3}$ and
$\omega = 1.8 \alpha^{1/3} A^{2/3}$. The spatial step $h$ of the
spiral far from the core will be $h = 10 \alpha^{-1/6}
A^{-1/3}$. Notice that a similar method was recently used to analyze
asymptotically the spiral waves in N-systems \cite{karma91,karma92}.

Comparing the results obtained above with Eq. (\ref{ab}), one can see
that in order for the solution in the form of the traveling front to
exist, one should have $A \gtrsim \alpha^{-1/2} \gg 1$. On the other
hand, for $A \gg \alpha^{-1/2}$ we have $\omega \gg 1$, so $\theta$
will not have enough time to decay behind the wave front. This means
that this kind of spiral wave may exist only at $A \sim
\alpha^{-1/2}$. Notice that according to Eq. (\ref{ca}) we have
$c_\infty \sim 1$ and $h \sim 1$ for these values of $A$, so the
formulas obtained above for the frequency of the spiral should only be
correct qualitatively.

These facts may seem rather surprising since they predict that the
spiral waves should exist only in a narrow range of the values of $A$
far from the excitability threshold $A = A_{bT} = 3.76 \alpha^{1/2}$,
which is obtained from Eq. (\ref{ab}) for $\eta_i = \eta_h = 1$
\cite{mo1:98}. Also, numerical simulations show that spiral waves
exist in a wide range of the values of $A$. What we will show below is
that the spiral waves actually exist for all values of $A_b < A
\lesssim \alpha^{-1/2}$.

The thing is that in addition to the spiral wave solution whose tip is
at rest, there may also be a solution whose tip moves along a circle
of some radius $R$, which can be large for $A \alpha^{1/2} \ll 1$. The
reason the front will not propagate inside the circle is that the tip
is right at the propagation threshold. This means that for $A \ll
\alpha^{-1/2}$ we have $\eta_i = \eta_i^b = 3.76 \alpha^{1/2} A^{-1}$
in the limit $\alpha \rightarrow 0$ (for $R \gg 1$ the corrections due
to curvature can be neglected), so the frequency $\omega$ is
determined by Eq. (\ref{slow}) with $\eta_i = \eta_i^b$. In
particular, for $\alpha^{1/2} \ll A \ll \alpha^{-1/2}$ we can expand
the exponential and obtain asymptotically
\begin{equation} \label{om}
\omega = 1.76 \alpha^{1/2} A.
\end{equation}
This equation shows that the value of $\omega$ lies in the range
$\alpha \lesssim \omega \lesssim 1$, as should be expected. For large
values of $R$ the speed of the front far away from the core should
only slightly exceed $c_{\mathrm min}$, so the step of the spiral will
be $h = 5.4 \alpha^{-1/2} A^{-1}$. Note, however, that because of the
closeness to the threshold point $\tilde{A} = \tilde{A}_b$ the
expansion in Eq. (\ref{dc}) is no longer justified and therefore the
BCF theory, as well as Eq. (\ref{ci}), is not applicable to the spiral
waves in this parameter range. This theory can be modified by noting
that close to $\tilde{A}_b$ we have approximately
\begin{equation} \label{c}
c = c_{\mathrm min} + b \left( \tilde{A} - \tilde{A}_b - {\tilde{A}_b
\over 2 c_{\mathrm min} } K \right)^{1/2},
\end{equation}
where $b$ is a constant and the tilde quantities in the right-hand
side are evaluated at $K = 0$. The analysis of Eqs. (\ref{travtheta})
and (\ref{traveta}) shows that $b = 1.4$. Observe that this equation
reduces to the form of Eq. (\ref{ci}) only very far from the core. 

Following \cite{burton51}, we rewrite Eq. (\ref{c}) for the steadily
rotating spiral in terms of the angle $\phi$ between the tangent
vector to the front and the radius vector as a function of the
distance $r$ to the origin. A straightforward calculation shows that
in these variables Eq. (\ref{c}) becomes
\begin{eqnarray} \label{kin}
{d \phi \over d r} = - {1 \over r} \tan\phi && \nonumber \\ && 
\hspace{-1cm} + {2 c_{\mathrm min} \over \tilde{A}_b b^2 \cos\phi}
\left[ b^2 (\tilde{A} - \tilde{A}_b) - (c_{\mathrm min} - \omega r
\cos\phi)^2 \right], \nonumber \\ \label{mod}
\end{eqnarray}
with the boundary conditions $\phi(+\infty) = \pi/2$ \cite{burton51}
and $\phi(R) = 0$. The latter condition says that at its tip the front
is normal to the circle along which it rotates, what follows from the
physical considerations. Since the front at $r = R$ is at its
propagation threshold, its normal velocity there should be equal to
$c_{\mathrm min}$. This, together with the boundary condition at $r =
R$ immediately gives us $R = c_{\mathrm min} / \omega$. Knowing the
value of $R$, one can then calculate $\tilde{A} - \tilde{A}_b$ and
find the value of $\omega$ for which it agrees with
Eq. (\ref{slow}). This will determine a unique value of
$\omega$. Numerical solution of Eq. (\ref{kin}) shows that for $\omega
\ll 1$ we have $\tilde{A} - \tilde{A}_b \ll 1$, so Eq. (\ref{om})
should indeed be recovered in the limit $\alpha \rightarrow 0$ with
$\alpha^{1/2} \ll A \ll \alpha^{-1/2}$, and the spiral wave solution
is an involute of a circle of radius $R$.

\begin{figure}
\centerline{\psfig{figure=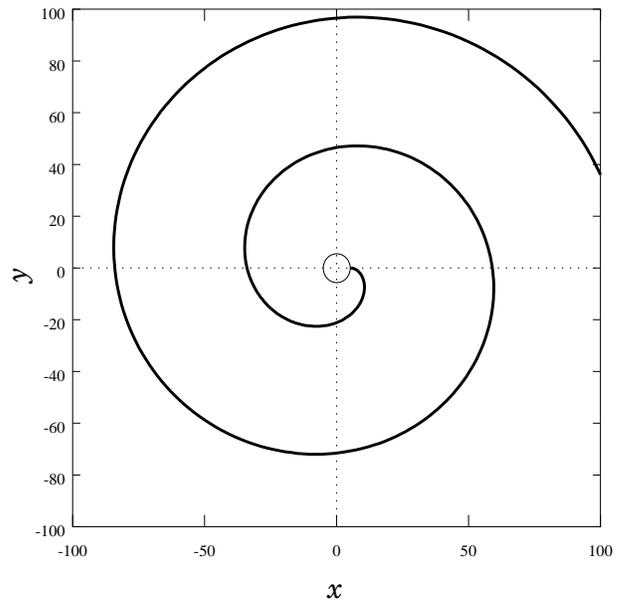,width=8cm}}
\caption{Steadily rotating spiral wave. Results of the numerical
solution of Eq. (\ref{mod}) with $\omega = 0.29$ and $\tilde{A} -
\tilde{A}_b = 0.34$. The circle shows the core of the spiral.}
\label{f2}
\end{figure}

The solution of Eq. (\ref{mod}) for $\omega = 0.29$ is presented in
Fig. \ref{f2}. For this value of $\omega$ we found $\tilde{A} -
\tilde{A}_b = 0.34$, which gives $\eta_i = 0.86$, within $4\%$ in
agreement with Eq. (\ref{slow}). Comparing these quantities with the
results of the numerical simulations of Eqs. (\ref{act}) and
(\ref{inh}) for $\alpha = 0.1$ and $A = 1.5$, in which this value of
$\omega$ was observed, we find that the value of $\eta_i$ agrees with
the predicted one within 3\% accuracy. The speed $c_\infty = 2.3$
obtained from Eq. (\ref{c}) also agrees with that observed numerically
within a few per cent accuracy.

This agreement is quite remarkable considering the fact that at these
parameters the spiral wave already underwent meandering
instability. In fact, according to our analysis, steady rotation of
the spiral requires fine-tuning of the value of $\eta_i$ at the tip of
the spiral. Note that the tip of the spiral is not described by the
interfacial equations derived above and thus is a rather singular
object capable of sudden movements on the smallest length scale. Thus,
it is natural to expect that the tip trajectory in a meandering spiral
may be rather abrupt. Notice that similar situation is observed in the
simulations of models of cardiac tissue (see, for example,
\cite{fenton98}). 

In conclusion, we developed a theory of spike spiral waves in
Gray-Scott model. Spike traveling waves are observed in a variety of
excitable systems including nerve and cardiac tissue. Even though we
performed an analysis of a concrete system, Eqs. (\ref{c}) and
(\ref{mod}) have general character and thus are expected to apply to
other $\Lambda$- and V-systems (see also \cite{ko:book,ko:ufn89}) and
other excitable systems of different nature in which spike traveling
waves are observed. Also, such waves can be expected in combustion
systems and Belousov-Zhabotinsky reaction in continuous flow
reactors. The thing is that although in Eqs. (\ref{act:gen}) and
(\ref{inh:gen}) describing these systems the activator nullcline may
formally be N-shaped, for typical parameters the value of
$\theta_{\mathrm max}$ may be several orders of magnitude greater than
$\theta_h$, so the system effectively behaves as a $\Lambda$- or
V-system. In particular, this is true for the models of systems with
uniformly generated combustion material \cite{ko:book,ko:ufn89} and
the two-parameter version of Oregonator \cite{tyson80}.

We would like to acknowledge computational support from Boston
University Center for Computational Science. One of us (C. B. M.) is
grateful to A. Karma for stimulating discussions.

\vspace{1cm}

\bibliography{../main}

\end{document}